\documentclass[a4]{IEEEtran}
\usepackage[pdftex]{graphicx}
\usepackage{epsfig}
\usepackage{amssymb}
\usepackage{amsmath}
\usepackage{amsfonts}
\usepackage{subfigure}
\usepackage{hyperref}
\begin{document}


%

\title{Side-Channel Oscilloscope}
\author{Sumanta Chaudhuri, Sylvain Guilley \\
Institut T\'EL\'ECOM / T\'EL\'ECOM Paris, CNRS -- LTCI (UMR 5141)\\
46 rue Barrault, 75\,634 PARIS Cedex 13, FRANCE. \\
}

\maketitle
\begin{abstract}
Side-Channel Analysis used for codebreaking could be used constructively as a probing
tool for internal gates in integrated circuits. This paper outlines basic methods
and mathematics for that purpose
\end{abstract}

\section{Introduction}
\label{sec:intro}
In recent times Side-Channel Attacks~\cite{kocher-dpa_and_related_attacks} have attracted a lot of attention from the information 
security community. These techniques are very similar to Spectroscopy(NMR) used over the years.
While in spectroscopy, patterns in the light spectrum are used to detect the presence of atoms and its environments
in an unknown substance, in a Side-Channel Attack the cryptanalyst looks for patterns in the power consumption or EM 
emission to detect the unknown key value. Apart from cryptanalysis, side-channel could be of great utility to an electrical 
engineer, as a probing tool more in line with spectroscopy used in physics. 
\section{Power Consumption Model}
\begin{figure*}
\centering
\includegraphics[width=1\textwidth]{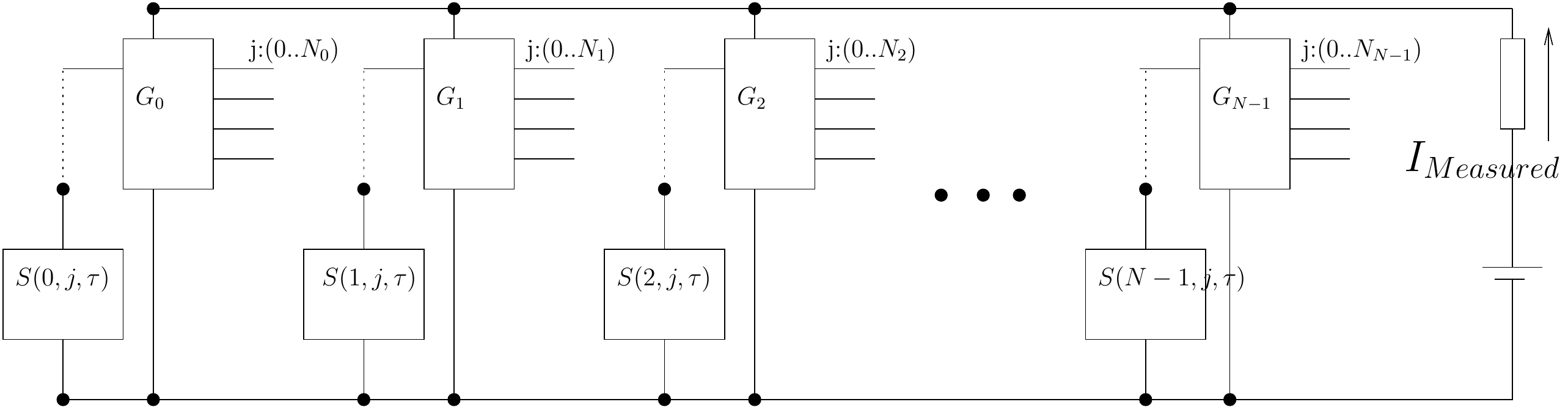}
\caption{Power Consumption Model.}
\label{fig:model}
\end{figure*}
In this article we consider the Power Traces $S(I_i,I_{i-1},t)$ of a combinatorial circuit. That is it has no memory. 
$I_i$ and $I_{i-1}$ are the $i^{th}$ and $(i-1)^{th}$ input to this block, and $t$ is the time sample of the measured 
current from power supply. Power is only consumed when the input goes through a transition. The power trace is recorded 
for a time duration [0,T] where the transition is applied at t=0 and and T is any time after that when there is no further 
change in  measured current. 
In this article we use a recursive model of power consumption. For purpose of illustration, we consider a block with N subblocks (see figure~\ref{fig:model}.

A sub-block with $N$ inputs is characterized by it's step current response as the input vectors undergoes a transition $i_j\rightarrow i_{k}$. Note that output loads
are considered to be part of this sub-block.
\begin{equation}
S^{i_j\rightarrow i_{k}}(t)= I_{measured}(t)
\end{equation}

and the sub-block is characterized by the set of all step responses corresponding to each transition.
\begin{displaymath}
\bigcup_{\substack{0<i_j<2^N-1 \\ 0<i_k<2^N-1}} S^{i_j\rightarrow i_{k}}(\tau)
\end{displaymath}

Now we view the DUT as recursively organized in various sub-blocks. At the topmost level the DUT consists of N such gates(see figure.~\ref{fig:model}) 
where $k_{th}$ gate has $N_k$ possible input transitions at it's input. That is, the length of the input transition alphabet to the $k_{th}$ gate is 
denoted as $N_k$. Furthermore 
\begin{itemize}
\item
$S(k,j,t)$ denotes the step current response $s(\tau)$ associated with $j_{th}$ 
transition of the $k_{th}$ gate. 
\item
$A(k,j,I_i,I_{i-1})$ denotes the $j_{th}$ transition on the $k_{th}$ block is activated, during the interval depending on $I_i$ and $I_{i-1}$
\end{itemize}

Next we normalize the traces to have a zero mean. So we will assume that $S(k,j,t)$ has a zero mean from now onwards.
We even redefine the activation function as:
\small
\begin{eqnarray*}
T(k,j,I_i,I_{i-1}) & = & 1 \quad\textrm{if input vector to the $k_{th}$ gate}{}
		\\
		& & {}\qquad\qquad \textrm{undergoes $j_{th}$ transition}\\
T(k,j,I_i,I_{i-1}) & = & -1 \quad\textrm{if input vector to the $k_{th}$ gate}{}
		\\
		& & {}\qquad\qquad \textrm{remains unchanged}\\
T(k,j,I_i,I_{i-1}) & = & 0 \quad \textrm{if $j> N_k$}\\
\end{eqnarray*}
\normalsize
Note that $A(k,,j,I_i,I_{i-1})$ can be written in terms of $T(k,,j,I_i,I_{i-1})$ as 
\begin{equation}
A(k,j,I_i,I_{i-1})=\frac{1}{2} \times (1+T(k,j,I_i,I_{i-1}))
\end{equation}

The advantage of this representation is that we can write, for M random input transitions
\begin{displaymath}
\sum_{i=0}^{M-1} T(k_1,j_1,I_i,I_{i-1}) \times T(k_2,j_2,I_i,I_{i-1}) = 0 \\
											\quad\textrm{if}  k_1 \neq k_2 \& j_1 \neq j_2
\end{displaymath}
\begin{displaymath}
\sum_{i=0}^{M-1} T(k_1,j_1,I_i,I_{i-1}) \times T(k_2,j_2,I_i,I_{i-1}) = M \\
										\quad	\textrm{if}  k_1 =  k_2 \& j_1 = j_2
\end{displaymath}
This is based on the assumption that all transitions are independent, and it closely follows the mathematical
definition of orthogonality over M random input transitions.

With this notation we can write one power trace for input vector transition from $I_i$ to $I_{i-1}$ as 
\begin{eqnarray}
\label{eq:1}
S(t)= \sum_{k=0}^{N-1} \sum_{j=0}^{N_k-1} A(k,j,I_i,I_{i-1}) \times S(k,j,n)\\
S(t)= \sum_{k=0}^{N-1} \sum_{j=0}^{N_k-1} \frac{1}{2} \times T(k,j,I_i,I_{i-1}) \times S(k,j,t)
\end{eqnarray}

\section{Post-Processing}
Now we want to find out the step current response associated with $p_{th}$ transition of the $q_{th}$ gate. For that purpose we 
apply M random transitions $<I_i,I_{i-1}>$ at the input which also includes transitions that will trigger the event $A(p,q,t)$ 
and multiply each trace by $T(p,q,I_i,I_{i-1})$
\small
\begin{eqnarray*}
s_{acc}(t) & = & \sum_{i=0}^{M-1} T(p,q,I_i,I_{i-1}) \times S(t) \\
s_{acc}(t) & = & \sum_{i=0}^{M-1} T(p,q,I_i,I_{i-1}) \times \sum_{k=0}^{N-1} \sum_{j=0}^{N_k-1} \frac{1}{2} \times {}
	\\
	& &{}(1+T(k,j,I_i,I_{i-1})) \times S(k,j,t) \\
s_{acc}(t) & = & \sum_{i=0}^{M-1} T(p,q,I_i,I_{i-1}) \times \sum_{k=0}^{N-1} \sum_{j=0}^{N_k-1} \frac{1}{2} \times S(k,j,t)+ {}
	\\
	& & {}  \sum_{i=0}^{M-1} \sum_{k=0}^{N-1} \sum_{j=0}^{N_k-1} (T(k,j,I_i,I_{i-1})) \times T(p,q,I_i,I_{i-1}) \times S(k,j,t) \\
s_{acc}(t) & = & \sum_{i=0}^{M-1} T(p,q,I_i,I_{i-1}) \times N \times N_k \times \frac{1}{2} E_{j,k}[S(k,j,t)]+ {}
	\\
	& & {}  \sum_{k=0}^{N-1} \sum_{j=0}^{N_k-1} S(k,j,t) \times  \frac{1}{2} \times {}\\
	& & {}  \sum_{i=0}^{M-1} (T(k,j,I_i,I_{i-1})) \times T(p,q,I_i,I_{i-1}) \\
s_{acc}(t) & = & \sum_{i=0}^{M-1} T(p,q,I_i,I_{i-1}) \times N \times N_k \times \frac{1}{2} E(S(t)) + \frac{M}{2} S(p,q,t)
\end{eqnarray*}
\normalsize

Since we preprocessed the traces$S(t)$ to have a zero mean , we can find out the step current response as
\begin{eqnarray}
s_{acc}(t) & = & \frac{M}{2} S(p,q,t) 
\end{eqnarray}
\vspace{1em}
\section{Recursive Refinement}
In the above section we outlined a method to find the current response associated with the $p_{th}$ transition of the $q_{th}$ gate.
This process can continued recursively for the $q_{th}$ gate until we have only a single net, in which case we can derive
the voltage waveform from the step current response using basic circuit behaviour.

In the above mentioned method, the orthogonality of $T(k,j,I_i,I_{i-1})$ functions play a pivotal role. Even in Template Side-Channel Attacks~\cite{Template}
A major step is to find orthogonal representation of the acquired traces.
To guarantee this orthogonality
we can divide a block recursively into two sub-blocks using minimum cut bisection, and finally arriving at the target transition.

\section{Conclusion}

In this article we illustrated a very preliminary outline of how power analysis techniques can be used for probing each single net 
behaviour in the circuit, thus acting as an oscilloscope even for physically inaccessible components. This will beneficial for modelling 
new technologies based on in-circuit measurement. The difference between Side-Channel Attacks(SCA)
and Side-Channel Oscilloscope(SCO) is that in SCA, the user does not have the full knowledge of the circuit. So the functions $T(k,j,I_i,I_{i-1})$ are only guesses.
In SCO, the user can calculate the activation functions, but for him the unknown is the response of the fabricated circuit.The major assumptions that we made
are that the transitions are orthogonal/independent, which may not be true for all circuits, however for some amount of interdependence
we still get a magnification for the target transition, and dependent transition current response  remains present as noise. 
This process can be further improved by imposing DFT rules, and using more complicated post processing techniques such as Principal Component Analysis
as used in Template Attacks~\cite{Template}.

\bibliographystyle{abbrv}
\bibliography{dac09}  
\end{document}